# 27 Free-space quantum key distribution

**Alberto Carrasco-Casado; Verónica Fernández; Natalia Denisenko**

Institute of Physical and Information Technologies (ITEFI)
Spanish National Research Council (CSIC)
Serrano 144, Madrid 28006, Spain
email: veronica.fernandez@iec.csic.es

## Abstract

Based on the firm laws of physics rather than unproven foundations of mathematical complexity, quantum cryptography provides a radically different solution for encryption and promises unconditional security. Quantum cryptography systems are typically built between two nodes connected to each other through fiber optic. This chapter focuses on quantum cryptography systems operating over free space optical channels as a cost-effective and license-free alternative to fiber optic counterparts. It provides an overview of the different parts of an experimental free-space quantum-communication link developed in the Spanish National Research Council (Madrid, Spain).

### 27.1    Introduction

Quantum cryptography and more specifically, Quantum Key Distribution (QKD), has become probably the most developed branch of quantum information technologies. QKD resolves the distribution of keys without relying on the computational difficulty of performing certain mathematical functions such as one-way functions, considered the cornerstone of public-key cryptography. Relying on physical principles, QKD guarantees the secure distribution of keys even against a quantum computer attack. Although a universal quantum computer is still far in the horizon, simplest versions that perform certain type of algorithms are emerging, and governments, companies and national security agencies are becoming increasingly interested in quantum computing research, with large amounts of funding being invested in this field.

Free-space and optical fiber are the most commonly used transmission channels of QKD systems. Free-space links can be easily transported to different locations if required, as opposed to optical-fiber links, which are usually underground and cannot be easily operated. City-range free-space QKD links could be of interest to organizations such as financial, governmental, and military institutions located within urban areas who wish to establish highly secure point-to-point links. If desired, these links could also be integrated into optical-fiber metropolitan networks and provide higher bandwidth at congested points affected by poor connectivity.

## 27.2     Quantum key distribution protocols

In this section, the most common QKD protocols are introduced: the BB84 protocol, which is also the basis for more sophisticated protocols, such as the SARG protocol, or the decoy-state protocols. A simplified version of the BB84 protocol, the B92 protocol, will also be explained, as it is the protocol used for the developed QKD system.

### 27.2.1 BB84 Protocol

The BB84 protocol uses a quantum channel with one important requirement: the capability of transmitting individual photons. Another channel is also needed, which can be any public line containing non-encrypted information, meaning that Eve can listen in without being detected (see Fig. 1).

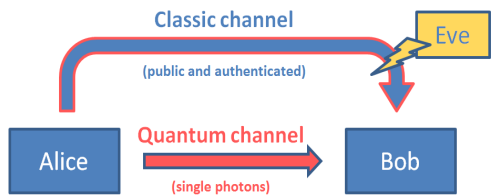

**Fig. 27.1.** Basic schematics of a quantum key distribution system.

Two conjugate variables, such as non-orthogonal polarization or phase basis sets are used in the BB84 protocol for encrypting information in individual photons. Being non orthogonal means that if a state is produced in a basis set and is measured using a different set, the result of the measurement will be necessarily random, as Heisenberg's Uncertainty Principle states. Conversely, using the same basis set will result in a correct measurement. These non-orthogonal basis set can be, in the case of polarization, a rectilinear base composed of a vertical (90˚) and horizontal linearly polarized state (0˚) and a diagonal base, consisting of linearly polarized light at 45˚ and -45˚. Each bit, '0' or '1', could be encrypted in two different polarization states, one of each base, i.e., the binary state '0' will be encrypted by 0˚ in the rectilinear basis and by 45˚ in the diagonal basis. Analogously, the binary state '1' will be encrypted by 90˚ in the rectilinear basis and by -45˚ in the diagonal one.

The protocol (Fig. 2) can be divided in two phases: the 'quantum' transmission and the 'public' discussion, depending on the channel being used. During the first phase, i.e., the quantum communication, Alice randomly chooses the encoding polarization base between 0˚, 90˚, 45˚ or -45˚ for each bit to be transmitted and keeps the information of the base she used for each bit. On the other end, Bob measures each bit received from Alice using the rectilinear and diagonal bases randomly. This rate of photons is known as the raw key. He also keeps a record of the results he obtained and the basis he used. Once the quantum communications is finished, the public discussion begins. In this phase, Bob tells Alice (through an authenticated but public channel) which bases he used for each measurement. Since they both chose their basis randomly, these will be the same half of the times. They keep these bits and discard the remaining ones, where they chose different basis. This photon rate is known as the sifted key.

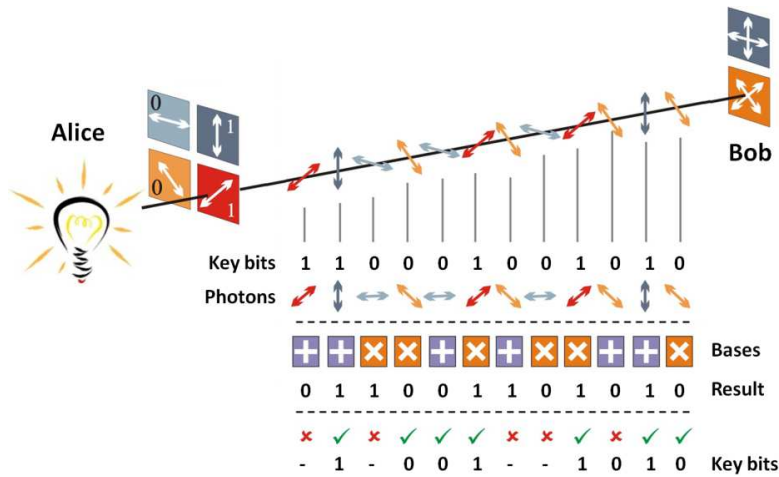

**Fig. 27.2.** BB84 protocol basic scheme.

Once Alice and Bob share a key, before using it, they need to verify that Eve did not intercept it. Alice and bob compare a fraction of the key that is selected in a random way. By comparing the fraction of wrong bits to the total fraction of received bits, they can calculate what is known as Quantum Bit Error Rate (QBER), which will tell them if the key has been compromised.

If Eve wanted to copy the key, she could in principle measure each bit and retransmit the outcome of this measurement. The problem is that Eve, as Bob, does not know the base in which Alice transmitted the photon. Therefore, on average, she will guess right 50% of the photons, correctly retransmitting the bits. From the other 50%, she will obtain a correct measurement in half of the cases. In total, 75% of the bits retransmitted by Eve will be correct, i.e., 25% are wrong. Therefore, this key will be sent to Bob. When Bob compares his basis with Alice ones and keeps only those that are the same, the error should be 0%, since he has kept only the correct measurements. If instead he finds a 25%, therefore Alice and Bob will know Eve was listening. The strength of the BB84 relies on the fact that Alice and Bob can verify if an eavesdropper was present in the channel by evaluating the error rate or QBER.

### *27.2.2. B92 Protocol*

In 1987, I. D. Ivanovic demonstrated that two non-orthogonal quantum states can be distinguished unambiguously at the cost of some loss. Based in this principle, C. H. Bennett proposed in 1992 a different version of the BB84 where only one state of each polarization basis set was used: the B92 protocol. The main difference with the BB84 protocol is that only two states are used instead of four. Besides, no basis reconciliation is needed; only the time slots where Bob measured an event must be shared with Alice to distill a common key. Bob still makes a 50% random basis choice but he measures the states unambiguously.

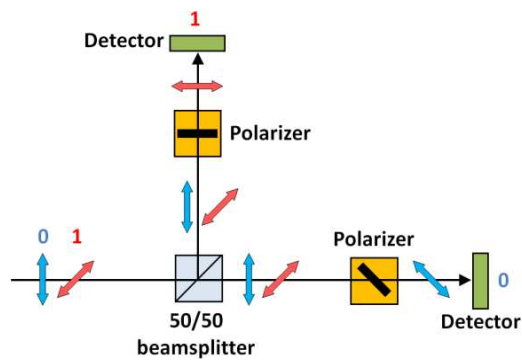

**Fig. 27.3.** The receiver in the B92 polarization-encoded protocol.

In order to comprehend the B92 protocol, the implementation of the polarization state detection in Bob must be well understood (Fig. 3). Alice encrypts her states in two linearly polarized states at a relative angle of 45˚. In Bob, as in the BB84 protocol, the basis selection is randomly chosen. A non-polarizing 50/50 beamsplitter models this basis selection, since a single photon will be transmitted or reflected with a probability of 50%, regardless of the incident polarization state. On the other hand, before the detectors, linear polarizers are placed at an opposite orientation to the state to be extinguished, i.e., in the channel where the '1's are desired to be detected, the linear polarizer is oriented to minimize the '0's; and analogously, in channel '0', the polarizer will be oriented against the polarization of the '1's. This ensures that only '1's are received in channel '1' and only '0's are received in channel '0'.

## 27.3    Free space as the 'quantum' channel

The transmission channel can be any medium capable of transporting the quantum states from the transmitter to the receiver of a quantum communication system. Its primary goal is preserving the quantum information encrypted in them and therefore it should have some important properties, such as low dispersion, low birefringence, etc.

In a realistic scenario, QKD will be implemented in metropolitan networks with a combination of optical-fiber and free-space links. While optical fiber is of enormous advantage due to its guiding properties, it sometimes lacks the flexibility of free-space links. These links are portable and can be relocated according to the needs of the network. Besides, there are locations where optical-fiber links are very difficult to install due to the geology of the terrain, such as cliffs, rivers, etc. Free-space links can be very useful in these scenarios. Besides, in optical networks, free-space links can serve as means of alleviating congestion in certain points of them, where high traffic is unavoidable due to the high number of clients connected to the network. Finally, it should be stressed that free-space QKD is of utmost importance for overcoming the distance limitation of quantum communication systems by incorporating satellite communication.

### *27.3.1 Transmission through the atmosphere*

The transmission $\tau$ of an optical beam through the atmosphere obeys Beers Law, according to equation (1).

$$\tau = \frac{I_R}{I_0} = e^{-\gamma x} \qquad (1)$$

In equation (1), $I_R$ is the detected intensity at a distance x from an initial Intensity $I_0$, and $\gamma$ is the attenuation coefficient. This attenuation coefficient is governed by two effects: the absorption and scattering coefficients ($\alpha$ and $\beta$, respectively) of molecules and aerosol particles present in the air, according to equation (2).

$$\gamma = \alpha_m + \alpha_a + \beta_m + \beta_a \qquad (2)$$

### *27.3.2 Scattering, absorption, and weather dependence*

### 27.3.2.1 Scattering

When light travels through any medium, its waveform gets distorted by the particles of this medium causing a loss of signal. It is a different process than absorption, since the energy is redistributed rather than absorbed. The intensity of the scattering depends on the size of the particles that light encounters. If r « $\lambda/2\pi$, the scattering is in the Rayleigh regime; for r ≈ $\lambda/2\pi$, the scattering is in the Mie regime; and for r » $\lambda/2\pi$, geometrical optics can be used to study the scattering process.

<u>Rayleigh scattering</u>. When light hits the bound electrons of an atom or molecule, a dipole is created which vibrates at the same frequency that the incident light. The dipole then reradiates the incident light in the form of a scattered wave. Rayleigh scattering is governed by the equation (3).

$$\sigma_S = \frac{f e^4 \lambda_0^4}{6\pi\varepsilon_0^2 m^2 c^4 \lambda^4} \qquad (3)$$

In equation (3), f is the oscillator strength, e is the charge of an electron, $\lambda$ is the wavelength corresponding to the atomic transition frequency $\nu = c/\lambda$, and c is the speed of light.

Fig. 4 shows how the $1/\lambda^4$ dependence implies that only shorter wavelengths will be affected by Rayleigh scattering, whereas for longer ($\lambda$ ~1500nm) wavelengths, this effect will be negligible.

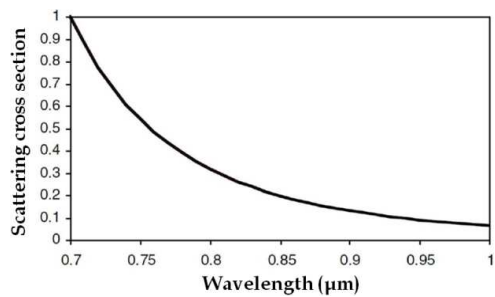

**Fig. 27.4.** Scattering cross section versus wavelength.

<u>Mie scattering</u>. In this case, the size of the molecules that scatter light are in the same scale as the wavelength. Therefore, fog, haze and aerosol particles are the main sources of scattering in this regime. In particular, fog is of special consideration since it can cause an extremely high loss of the signal. A simplified empirical formula is given by equation (4).

$$\gamma = \frac{3.91}{V}\left(\frac{\lambda}{550}\right)^{-\delta} \tag{4}$$

In equation (4), $\delta = 0.585(V)^{\frac{1}{3}}$ for $V < 6$ km, $\delta = 1.6$ for $V > 50$ km and $= 1.3$ for $V < 50$ km, V corresponding to the visibility of the propagation path. This formula however has been questioned by the scientific community as it does not reproduce exactly the empirical observations. More precise numerical simulations of the exact Mie scattering formula conclude that the wavelength dependence in the near infrared range is initially thought and that fog is the cause of the highest transmission loss. It must also be said that collecting empirical data is difficult, as the location, time, wind velocity and relative humidity can affect aerosol distribution.

## 27.3.2.2 Absorption

The absorption coefficient α of an absorbent particle depends on its size and on the distribution of these particles along the beam path, according to the equation (5).

$$\alpha = \sigma N \tag{5}$$

In equation (5), σ is the cross section and N the concentration of the absorbing particles. There are two main types of absorbers in the atmosphere: molecular and aerosol absorbers, depending on the type of particle that influences in beam transmission.

<u>Molecular absorption.</u> It occurs when the beam particles interact with the molecules present in the atmosphere such as $N_2$, $O_2$, $H_2$, $H_2$, $CO_2$, $O_3$, etc. If the incident wave corresponds with one of the resonant frequency of the present gases, molecular absorption takes place, resulting in the so-called blocking windows of the typical atmospheric absorption spectra (see Figure 5). In the infrared region the main molecular absorbers are water vapor, carbon dioxide and ozone, being the most affecting absorber water vapor for the near infrared spectrum (up to 3 μm).

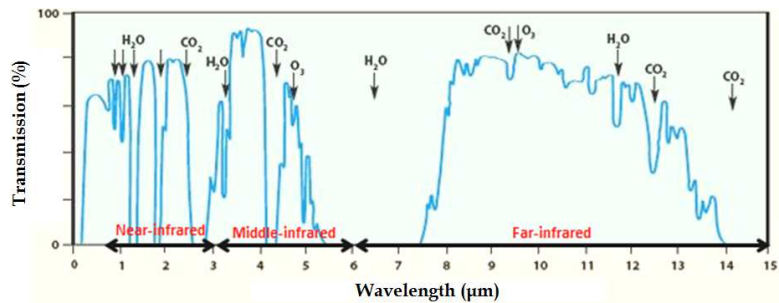

**Fig. 27.5.** Atmospheric transmission versus wavelength for the UV, visible and infrared bands.

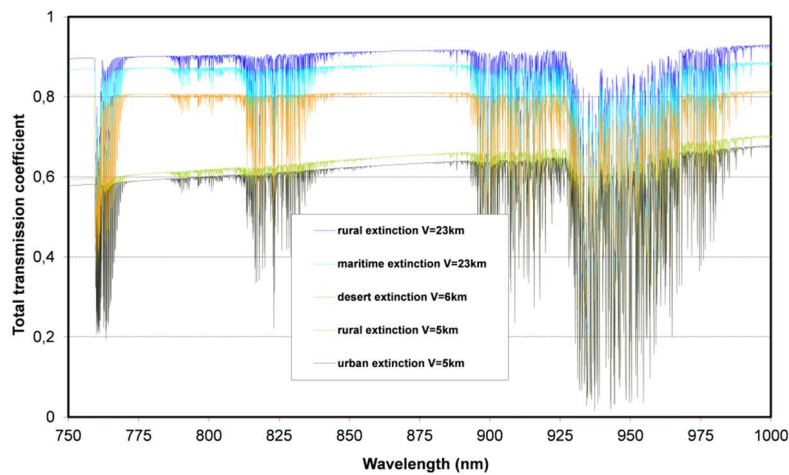

**Fig. 27.6.** Atmospheric transmission for a 1-km path, at an altitude of 1 m for several locations and visibility ranges.

<u>Aerosol absorption</u>. Aerosols are fine liquid or solid particles in suspension in the atmosphere of size ranging between $10^{-2}$ and $10^2$ μm. Their origin is diverse and depends on the geographic location, such as droplets of water (drizzle, fog, foam) along with salt crystals in maritime regions; or human-made aerosols in urban areas, such as sulfates from coal power plants or black and organic carbon from motor vehicle emissions. The aerosol concentration also depends on the time of the day and meteorological conditions and influence atmospheric transmission differently depending on their concentration, size and chemistry. This gives rise to the term visibility, which defines the line of sight of the optical link. Fig. 6 shows atmospheric transmission depending on the location and visibility of the link.

### 27.3.2.3 Weather dependence

The weather conditions and visibility strong influences atmospheric transmission. In the case of light to moderate rain, the diameter of the rain droplets are between 200-2000 μm, which is considerably larger than the wavelengths involved in FSO and thus atmospheric transmission is not dramatically impacted, as in the case of heavy rain (see Fig. 7). Snow conditions are not so relevant, since the ice particles tend to be larger than rain droplets and thus their influence is lower. However, fog is the weather condition that most jeopardizes FSO transmission as the size of moisture particles are of the same size as near-infrared wavelengths. The term fog is used when the visibility is lower than 2000 m. For longer visibilities, the weather conditions are normally called hazy. Depending on the visibility range, fog can be considered thin (for visibilities between 1.9 km and 2 km), light (770 m-1 km), moderate (500 m), thick (200 m) and dense fog (50 m).

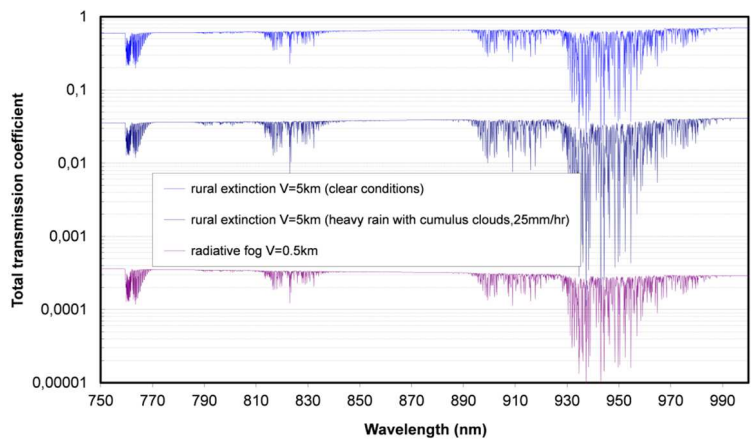

**Fig. 27.7.** Atmospheric transmission for a 1-km path, at an altitude of 1 m for several weather conditions.

However, most of these data proceeds from measurement taken at airports, which are located usually at a certain distance from cities, and are usually averaged over many years. Therefore, local effects on fog or the effect of altitude on fog conditions might not be taken into account, thus the predication of the exact effect of fog is complex and one of the bigger challenges in FSO communications.

### *27.3.3 Atmospheric turbulence*

The origin of atmospheric turbulence comes from temperature variations of the atmospheric air. When the temperature rises, the density of air reduces, and its molecules go up to higher layers. Cold air falls to lower layers because of its higher density. This way, turbulent movements of masses of air emerge. These movements end up causing variations in the refractive index, creating random fluctuations in time and space. This way, electromagnetic signals that propagate through the atmosphere suffer from random phase and amplitude modifications.

These turbulent variations were synthesized by A. L. Cauchy in an expression, later experimentally verified, which E. Lorenz extended to take into account the effects of humidity, resulting in the Cauchy-Lorenz formula, as in equation (6). In this equation, n is the refractive index, $\lambda$ is the wavelength in $\mu$m, p and v are the dry air pressure and water vapor pressure in mb, respectively, and T is the temperature in K.

$$n-1 = \left( \frac{77,6 \cdot 10^{-6}}{T} \right) \left( 1 + \frac{7,52 \cdot 10^{-3}}{\lambda^2} \right) \left( p + 4810 \frac{v}{T} \right) \qquad (6)$$

It is common to consider that the refractive index depends on the temperature variations, much more than on the pressure variations. These temperature variations show a frequency behavior proportional to $k^{-5/3}$ (being k the wave number $k = 2\pi/\lambda$) according to the classic Kolmogorov turbulence model. This dependence is expressed using a structure function, which averages the fluctuation over an interval r, as in equation, where T(x) is the temperature in a given point and T(x+r) is the temperature in a point at a distance r from the first one.

$$D_T\left( r \right) = \left\langle \left( T\left( x+r \right) - T\left( x \right) \right)^2 \right\rangle \qquad (7)$$

If the distance r between the two points is small (smaller than the inner scale $l_0$, related to the size of the smaller eddies), the structure function turns into equation (8), where $C_T^2$ is the structure parameter of temperature, which has an equivalent for the refractive index in the equation (9).

$$D_T\left( r \right) = C_T^2 r^{2/3} \qquad (8)$$

$$D_n(r) = C_n^2 r^{2/3} \qquad (9)$$

The structure parameter of refractive index $C_n^2$ from equation (9) describes the intensity of the fluctuations of the refractive index of atmospheric air. This parameter is always changing depending on the atmospheric conditions, on the height, on the location, etc. and its value can range over an order of magnitude within a few meters. Therefore, an average value is usually used, and it gives an estimation on the turbulent regime: weak ($C_n^2 \leq 10^{-17}$ m$^{-2/3}$), moderate ($10^{-17}$ m$^{-2/3} \leq C_n^2 \leq 10^{-13}$ m$^{-2/3}$) and strong regime ($C_n^2 \geq 10^{-13}$ m$^{-2/3}$).

## 27.4 Design of the transmitter: Alice

In this section, many considerations for designing the transmitter of a QKD system are reviewed, such as the transmission wavelength, the type of optical source, the optimization of relevant parameters such as the transmitting apertures, etc. Moreover, a description of the implemented free-space QKD transmitter referenced in [1] will be shown.

### 27.4.1 Choice of wavelength and source for the transmitter

The selection of the wavelength for a free-space QKD system is based on two main considerations: the transmission of this wavelength through the atmosphere and its detection efficiency at the receiver.

Two high-transmission spectral windows are usually employed in free-space QKD systems, which correspond to the wavelengths of $\lambda \sim 850$ nm and $\lambda \sim 1550$ nm. A third wavelength has also been used ($\lambda \sim 650$ nm) due to the ease of operation at a visible wavelength. Since the spectral regions of $\lambda \sim 850$ nm and $\lambda \sim 1550$ nm correspond to the first and third optical-fiber transmission windows, commonly employed by the telecommunications industry, off-the-shelf components at these wavelengths are readily available. In QKD systems, optical sources capable of high-speed modulation are also desirable, since they enable high-speed key transmission. This is of special interest due to the impossibility of optical amplification of the quantum states, albeit good progress is being made in the field of quantum repeaters [2]. Regarding operation frequency, laser diodes capable of high-speed modulation (Gbits/s) are available for λ ~ 850 nm and λ ~ 1550 nm, whereas they are uncommon at λ ~ 650 nm.

While λ~1550nm is, in terms of absorption in urban zones, the least attenuated in air (see Fig. 8), detection technology based in InGaAs/InP is still less advanced than silicon, outperformed in parameters like dark counts and detection efficiency. Although it must be stressed that great progress is being made in the field, regarding the operation frequency and GHz clock rates [3]. New single-photon detectors based in the superconducting effect [4] are promising candidates too, and efficiencies of more than 90% have recently been achieved at λ~1550nm [5]. However the need of cryogenic temperatures in the range of typically 0.1 mK-2 K is still their main disadvantage.

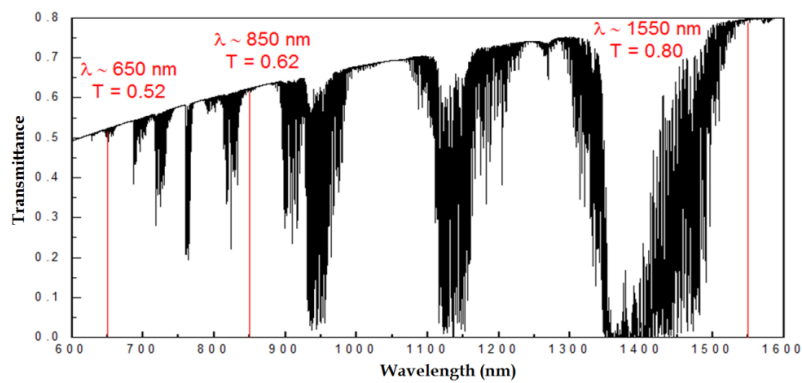

**Fig. 27.8.** Atmospheric transmission versus wavelength for the UV, visible and infrared bands for urban aerosol and a visibility of 5 km.

### 27.4.2 Optical configuration of the transmitter

The transmitter (see Fig. 9) must send out single photons carrying information encrypted in their polarization state. In order to emit individual photons, the laser source is heavily attenuated, thus emitting the so-called weak coherent pulses, which follow a Poisson distribution. This means that the probability of emitting a photon within a given pulse is always nonzero and therefore Alice's pulses must be heavily attenuated to decrease this probability to a sufficiently low value. In this case, the mean number of photons per pulse was μ ~ 0.1.

To achieve high transmission rates, Alice uses a fast gigahertz pulse pattern generator in conjunction with vertical-cavity surface emitting lasers controlled by high-bandwidth drivers, operating at several Gbit/s. Both VCSELs emit at a central wavelength of 848 nm, with an FWHM spectral bandwidth of 45 pm. The driver of each VCSEL contains an automatic power control circuitry, which maintains a constant optical power output and ensures a stable wavelength emission over changes in the temperature and other laser properties. Spatial modal filtering is achieved by coupling into a single-mode optical fiber at $\lambda \sim 850$ nm, ensuring that only one spatial mode is propagated. Two fiber-coupled attenuators then reduce the average number of photons per pulse down to $\mu \sim 0.1$, guaranteeing that only 0.5% of the emitted pulses contain more than one photon per pulse. Higher values of $\mu$ can be used if a decoy-state protocol is implemented, as reported in [6]. The optical output from the attenuators is then polarized and collimated to generate the two polarization states required by the B92 protocol. These states are combined by a non-polarizing beam splitter cube and a broadband pellicle beamsplitter and are redirected to a Kepler beam expander. The output beam from Alice has a diameter of ~40 mm.

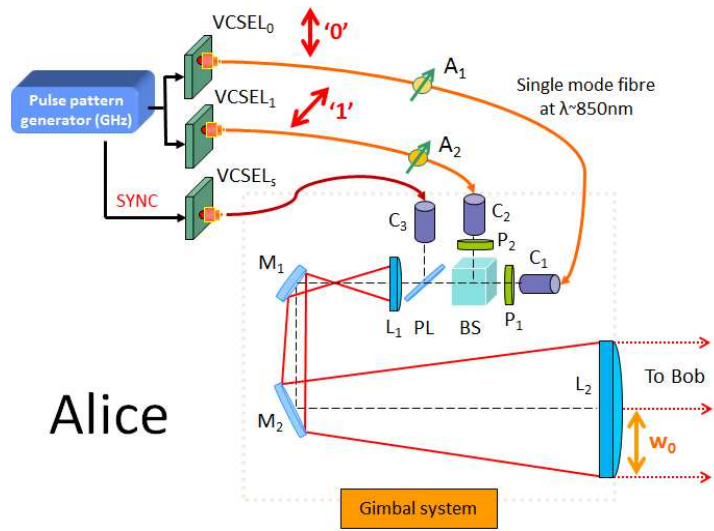

**Fig. 27.9.** Diagram of the transmitter: Alice. $VCSEL_0$, $VCSEL_1$ and $VCSEL_S$ are three vertical-cavity surface-emitting lasers; the first two emit at $\lambda$=850 nm to generate the quantum signal; and the third at $\lambda$≈1550 nm for the timing synchronization. These three lasers are combined by a non-polarizing beam splitter (BS) and a broadband pellicle beam splitter (PL); $A_0$ and $A_1$ are two optical-fiber attenuators; $C_0$, $C_1$ and $C_S$ are three fibre-coupled collimators; $P_0$ and $P_1$ are two high extinction-ratio polarizers; $L_1$ and $L_2$ are two achromatic doublet lenses which form the emitter's output telescope; and $M_1$ and $M_2$ are two high reflectivity mirrors.

Alice's optics was mounted on a gimbal system (see Fig. 10) capable of generating the tip-tilt movements necessary to be aligned with the receiver achieving a high precision (0.00025° for the vertical axis and 0.001° for the horizontal axis). The whole system was then sustained by a sturdy tripod. The transmitter was also isolated from the background light that couples through Alice's mirrors into Bob's detectors. Black-out material and a long tube was installed on Alice's objective lens ($L_2$) to decrease the field of view and thus the background noise.

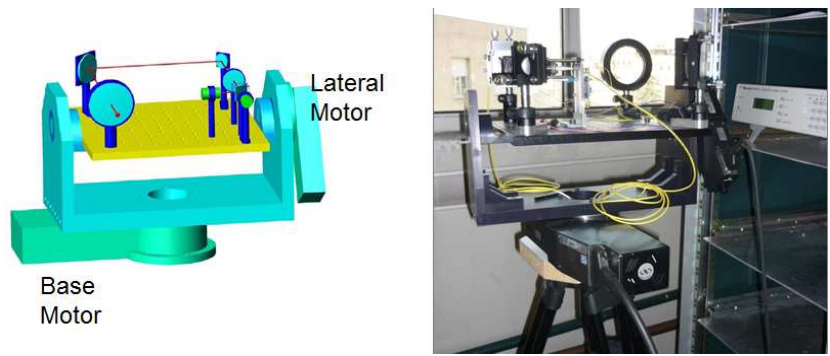

**Fig. 27.10.** CAD representation (left) and picture (right) of Alice's gimbal platform. Two motors control Alice' vertical and horizontal movements with high precision.

### 27.4.3 Temporal synchronization

The synchronization of the transmitter and the receiver is achieved by the transmission of periodic bright pulses at a different wavelength than that the quantum signals so that no interference or coupling takes place between them. The frequency of this signal is a sub-multiple value of the clock frequency and thus synchronous with the quantum signals, which avoids time delays between both signals. Fig. 11 shows a schematic of the synchronization process.

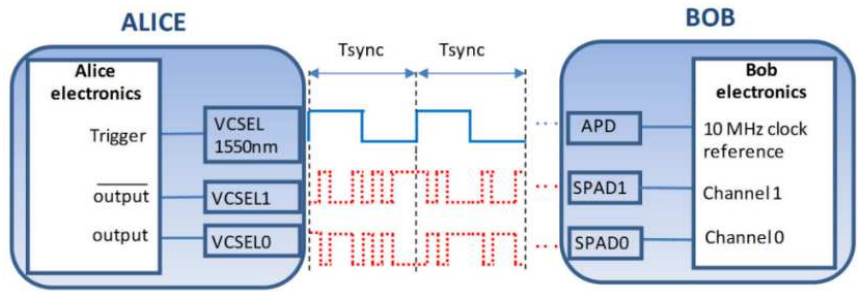

**Fig. 27.11.** Schematic showing the timing synchronisation between Alice and Bob. Alice transmits a periodic bright pulse at 1550 nm synchronous with the signal carrying the key at a subperiod Tsync of the original clock frequency from Alice's quantum signal. The synchronisation signal is detected at the receiver by an avalanche photodetector (APD) and serves as the reference clock for the timestamp card in Bob.

The wavelength for the synchronization was chosen to be $\lambda \sim 1550$ nm, since it coincides with an optical transmission band (see Fig. 8) and it is a standard telecommunication window, meaning that off-the shelf components are readily available at this wavelength. It is also far from $\lambda \sim 850$ nm, facilitating spectral discrimination in the receiver.

## 27.5 Design of the receiver: Bob

In this section, the critical parameters for designing the receiver of the free-space QKD system are reviewed. The optical setup is described in detail, along with the telescope and technical considerations, such as the correction of the polarization state and the detection technology of single photons.

### 27.5.1 Optical setup of the receiver

The fraction of photons from Alice that reach the receiver is collected by a Schmidt-Cassegrain Meade LX200 telescope, which focuses the beam along the optical axis and is then coupled into a 62.5 μm optical fiber with two achromatic 30 mm-focal distance lenses. The dichroic filter reflects the 1500 nm beam, which is sent to a 200 μm core diameter optical fiber and connected to an amplified photodetector.

This output is then connected to the 10 MHz external clock of the time interval analyzer (TIA) in order to acquire the same timing basis as Alice. The TIA used was the GT658 from GuideTech, which offers a bandwidth of 400 MHz, and a temporal resolution of 75 ps. Its purpose is the registration of the temporal events of the arrival times of the 850 nm photons. The arrival times of each channel were acquired through a piece of software programmed in LabView that could interface with the timestamp card. A schematic of Bob is shown in Fig. 12.

The 850 nm signal is splitted from the 1550 nm signal by a dichroic filter, and a narrow band-pass filter further eliminates other wavelengths from the sun spectrum. There are different technologies for detecting single photons, being the most commonly used avalanche detectors in Geiger mode. These detectors are avalanche photodiodes biased above the breakdown voltage, which means that a single photon is capable of initiating a self-sustaining avalanche. It is a mature technology, with high quantum efficiencies, few dark counts, high temporal resolution and commercially available. Besides, detection technology based in Silicon does not need cryogenic temperatures to operate as it is the case for Germanium or InGaAs for lowering the dark count levels. The SPADs used in the QKD system were SPCM-12-AQR from Perkin Elmer, which can detect 15 million counts per second according to specifications, with detection efficiencies around 45% and 500 counts per second. These detectors were characterized, showing an average detection efficiency of 30% and a dark count level of 300 counts per second.

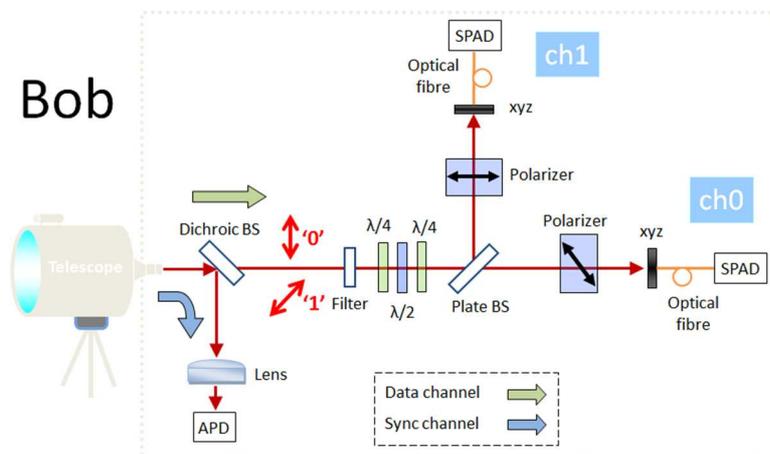

**Fig. 27.12.** Diagram of the receiver: Bob. In this diagram, λ/2 is a half waveplate; DM is a dichroic filter; L is an achromatic doublet lens, xyz is a three-axis positioner; APD is an avalanche photodetector, IF is an interference filter; BS is a beam splitter; λ/4 is a quarter waveplate; $P_0$ and $P_1$ are linear polarizers; SPAD are single photon avalanche detectors.

Regarding the optical detection of the polarized states, the 850 nm photons were randomly steered to each of the two detection basis of the B92 protocol through a 50/50 non-polarizing beamsplitter. In the reflecting path, the polarizer placed before the detector is oriented horizontally to block the vertical state and against the diagonal state, i.e., at -45°, in the transmitting path. This scheme ensures the deterministic discrimination of the two non-orthogonal states at the cost of incurring a loss of 75% of the photons: 50% are lost at the beamsplitter, which is what selects randomly the basis, so half of the states choose the wrong path and are blocked by the polarizer; and the other 50% choose the correct path but are blocked by the polarizers since they are oriented at an angle of 45° from their transmitting axis.

### 27.5.2 Single photon detection

Silicon single-photon avalanche detectors (SPADs) are widely used as single photon detectors for QKD systems due to their high efficiency, low noise, ease of operation and commercial availability. A SPAD is an avalanche photodiode (APD) biased at a voltage known as the breakdown voltage. The phenomenon called 'avalanche breakdown' (see Fig. 13) takes place at the voltage necessary for an electron-hole pair (generated by the absorption of a single photon) to gain sufficient kinetic energy (due to the existence of a high electric field across the device) to release further electron-hole pairs via the process of impact ionization. The process continues until a macroscopic current is generated. The single-photon detection regime is thus achieved when the APD is operated above breakdown (Geiger mode) and in this mode the APD is called single-photon avalanche diode (SPAD).

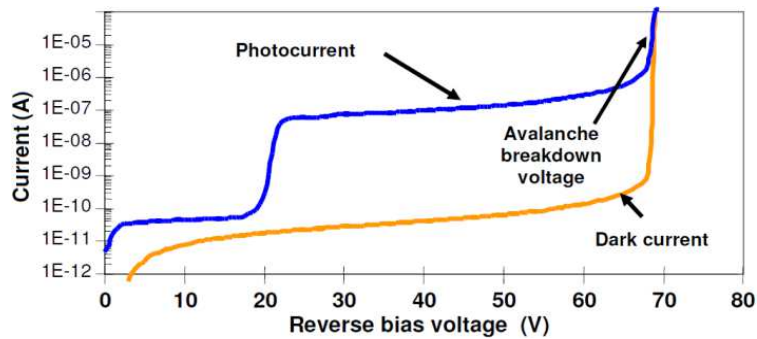

**Fig. 27.13.** Typical behavior of the photocurrent and dark current of an InGaAs APD at λ ~ 1300 nm fabricated by Fujitsu.

The SPADs used in the studied QKD system were commercially-available SPCM AQR 12 by PerkinElmer. Finally, recording the time arrival of each photon from the quantum signal detected by Bob was performed by a high precision time stamping card (GT658PCI from GuideTech) with a timing resolution of 75 ps.

## 27.6        Results of the QKD system

In the previous sections, the design and implementation of a high-speed free-space quantum key distribution system, conceived to operate in an urban environment, has been presented. This design has been carried out taking special care of the operating conditions of metropolitan networks, especially the non-stop 24/7 service. This implies being able to operate both under strong atmospheric turbulence and daylight background noise. In this final section, the QKD system is applied to a realistic environment to extract some results in terms of the quantum protocol parameters.

### 27.6.1 300-meter link experiment

The QKD system described in the previous sections was tested in the facilities of CSIC (Spanish National Research Council), in the city center of Madrid (Spain) for a 300-meter link, shown in Fig. 14. The system was characterized by performing measurements of the Quantum Bit Error Rate (QBER), the Secret Key Rate (SKR) and the background noise rate. These measurements were performed during the day, with high solar radiation, between spring and summer.

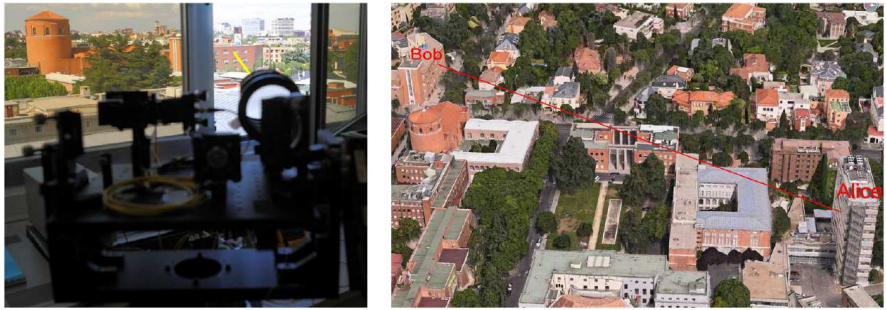

**Fig. 27.14.** 300-meter QKD link between Alice and Bob in CSIC facilities in Madrid (Spain).

The results of a 24-hour test of the QKD system are shown in the Fig. 15. There is an inverse relation between QBER and SKR, as the latter is used to calculate the former in the following way. The sifted key is generated by counting the events triggered by received photons as a 0 or a 1. These events are registered separately in two different detectors and each received photon is associated with a time tag (note that there is no raw key in B92, as there is no base reconciliation). At the beginning of the transmission, a short sequence, known by Alice and Bob, is transmitted in order to compute the QBER. Using this sample to count the number of errors (in which Alice and Bob do not agree) related to the total number of received bits, the QBER can be calculated. If the QBER is higher than a certain threshold, which depends on the protocol and the implementation (8% in the system described in the previous sections), the transmission is not considered to be secure. Lastly, the SKR is calculated from the sifted key rate and the QBER. In a real link, this step would go together with error correction and privacy amplification. In order to take this into account, a speed loss and the possible eavesdropping attacks, the key rate was estimated assuming cascade error correction algorithm and a worst-case scenario in which the two known attacks to B92 protocol (Unambiguous State Discrimination and Photon Number Splitting Attack) are performed simultaneously.

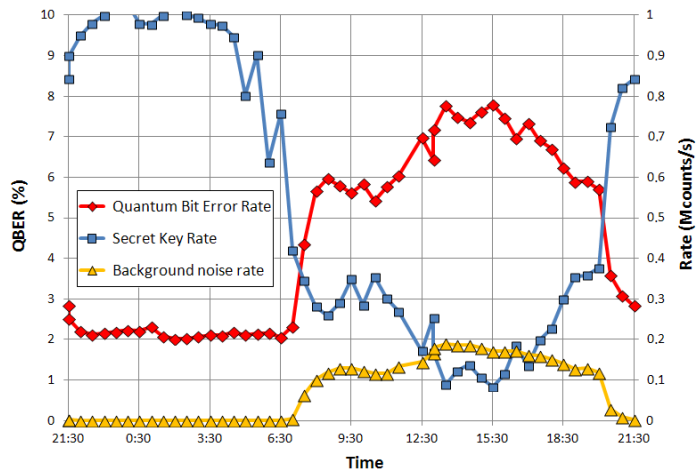

**Fig. 27.15.** Quantum Bit Error Rate (QBER), Secret Key Rate (SKR) and background noise rate in the 24-hour 300-meter link experiment.